\documentstyle[twoside,fleqn,espcrc2,epsf,eqnfix]{article}

\newcommand{\be} {\begin{equation}}
\newcommand{\ee} {\end{equation}}
\newcommand{\bdm} {\begin{displaymath}}
\newcommand{\edm} {\end{displaymath}}
\newcommand{\bc} {\begin{center}}
\newcommand{\ec} {\end{center}}
\newcommand{\beqa} {\begin{eqnarray}}
\newcommand{\eeqa} {\end{eqnarray}}

\newcommand{\bear}{\begin{eqnarray}}
\newcommand{\ear}{\end{eqnarray}}
\newcommand{\bea}{\begin{eqnarray*}}
\newcommand{\ea}{\end{eqnarray*}}

\def\pmb#1{\setbox0=\hbox{#1}
\kern.05em\copy0\kern-\wd0 \kern-.025em\raise.0433em\box0 }

\def\ell{l}

\def\P{I\!\!P}

\def\P{I\!\!P}
\def\pd{\partial}

\font\fiverm=cmr5
\font\sevenrm=cmr7

\newcommand{\AmS}{{\protect\the\textfont2
   A\kern-.1667em\lower.5ex\hbox{M}\kern-.125emS}}

\title{The proton's gluon structure}

\author{P V Landshoff%
\address{Centre for Mathematical Sciences \\
         Cambridge CB2 0WA}%
         \thanks{pvl@damtp.cam.ac.uk}}

\begin{document}

\begin{abstract}
The proton's gluon structure function at small $x$
is larger than nowadays is commonly believed. 
\end{abstract}

\maketitle

\begin{figure}[p]
\bc
\epsfxsize=0.8\hsize\epsfbox[50 50 400 300]{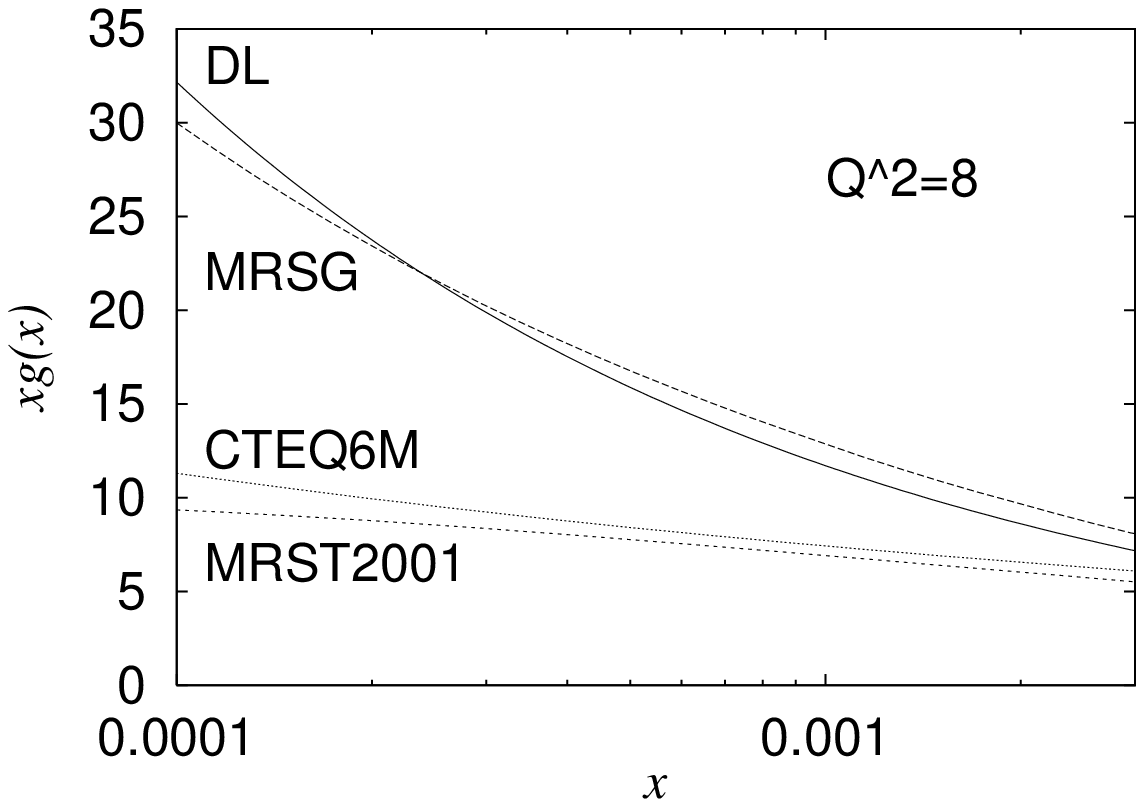}
\ec
\vskip -15truemm
\caption{Gluon structure functions from http://cpt19.dur.ac.uk/hepdata/pdf3.html and reference 1}
\label{GLUONSF}
\vskip 8truemm
\begin{center}
\epsfxsize=\hsize\epsfbox[100 300 410 760]{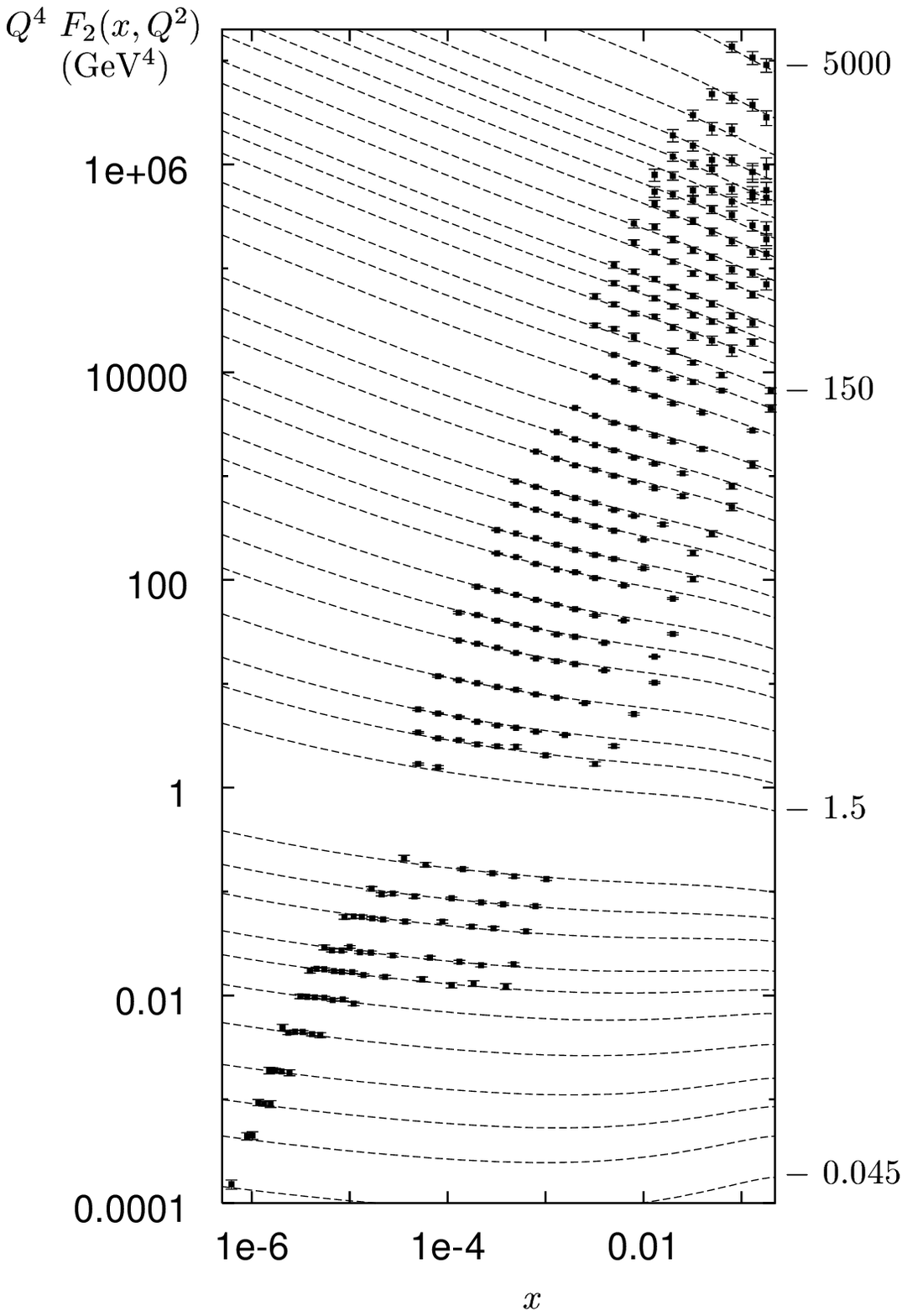}
\end{center}
\vskip -12truemm
\caption{Regge fit to ZEUS and H1 data for $F_2(x,Q^2)$ for $Q^2$ between 0.045
and 5000 GeV$^2$. The parameters were fixed using only data for
$x<0.001$ and therefore $Q^2\leq 35$ GeV$^2$.}
\label{F2}
\end{figure}
\section{Introduction}

We do not understand perturbative QCD at small $x$. In particular, as I
will explain, we do not understand how to apply DGLAP evolution there.
However, if we combine it with Regge theory and use an important message
from the HERA data for the charm structure function $F_2^c(x,Q^2)$, it
is possible\cite{dl} reliably to extract the gluon structure function $g(x,Q^2)$
at small $x$. It turns out to be larger than nowadays is commonly believed.
This is seen in figure 1. The most recent CTEQ and MRST structure 
functions\cite{mrst,cteq}
agree well with each other and with those extracted by the two HERA
experiments\cite{nagano} because they all use similar procedures; 
however, Donnachie
and I believe that the old MRSG structure function is nearer the truth.

\section{Regge theory -- the two pomerons}

When one tries to fit data, it is usually sensible to start with
the simplest assumptions and then refine them later. In its simplest
form, Regge theory leads to fixed powers of $x$ at small $x$, and it turns
out that two terms are enough: 
\be
F_2(x,Q^2)\sim f_0(Q^2)x^{-\epsilon_0}+f_1(Q^2)x^{-\epsilon_1}
\label{regge}
\ee
The second term corresponds to soft-pomeron exchange, with $\epsilon_1\approx
0.08$ determined from soft reactions. The data need a term that rises
more rapidly at small $x$; one needs $\epsilon_0\approx 0.4$. By fitting
the data at each $Q^2$, Donnachie and I found\cite{twopom} that
a successful and economical parametrisation of the coefficent functions
is provided by
$$
f_0(Q^2)=A_0 (Q^2)^{1+\epsilon_0}/(1+Q^2/Q_0^2)^{1+\epsilon_0/2}~~~~~~~~~~
$$
\begin{equation}
f_1(Q^2)=A_1 (Q^2)^{1+\epsilon_1}/(1+Q^2/Q_1^2)^{1+\epsilon_1}
\label{pheno}
\ee
with $Q_0\approx 3\hbox{ GeV}$ and $Q_1\approx 0.8\hbox{ GeV}$.
To make the fit, we used real-photon data and DIS data with $x\le 0.001$,
so that $Q^2$ ranges from 0.045 to 35 GeV$^2$. If we then simply multiply
the resulting form (\ref{regge}) by $(1-x)^7$, as is suggested by
the dimensional counting rules\cite{BF73b,MMT73}, it agrees quite well
with the HERA data even beyond $x=0.1$ and up to $Q^2=5000$ GeV$^2$.
This is shown in figure \ref{F2}. Note that this factor $(1-x)^7$ should not
be taken too seriously; it is much too simple. 
\begin{figure}
\bc
\epsfxsize=\hsize\epsfbox[75 380 370 770]{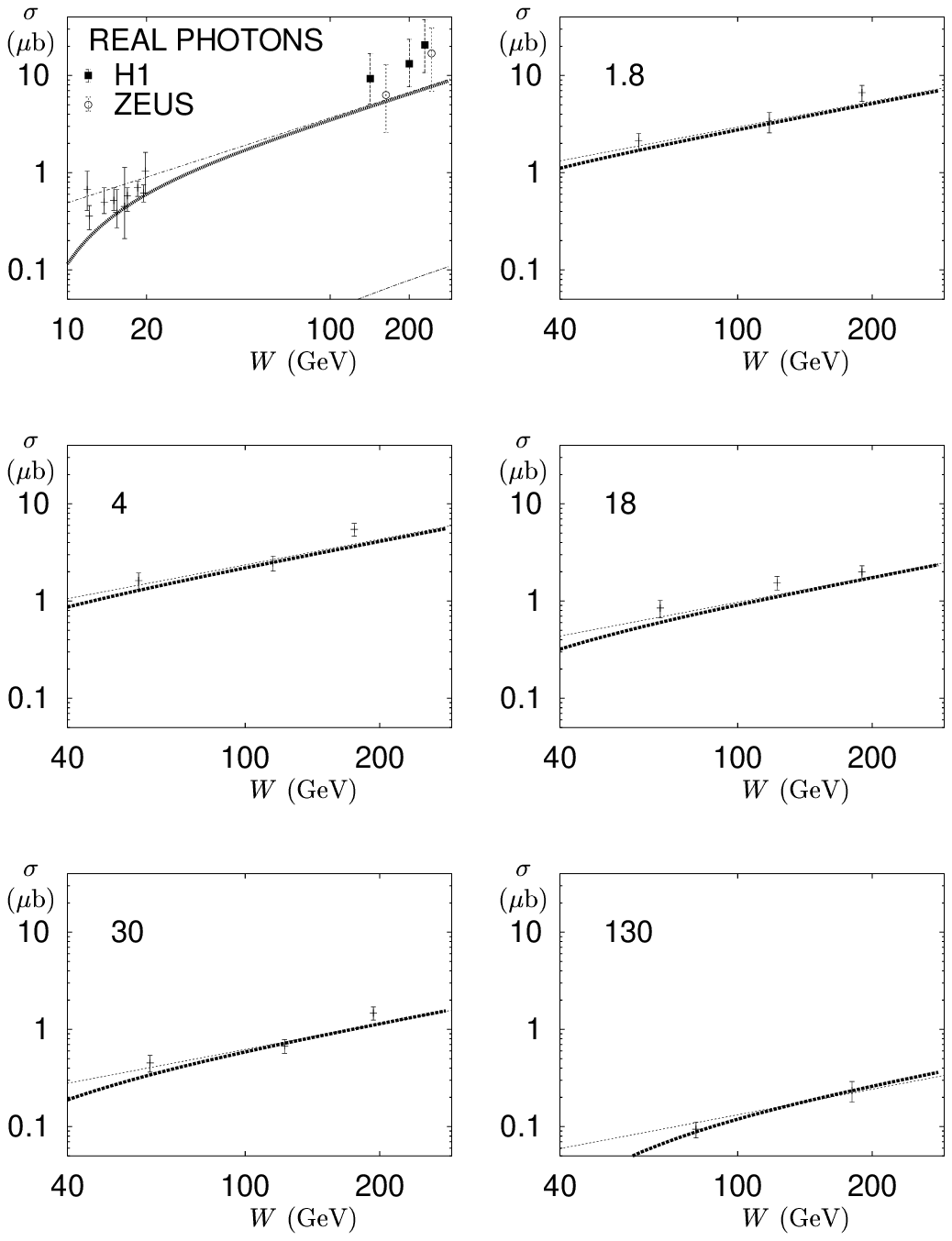}
\ec
\vskip -12truemm
\caption{Data\cite{zeusc} for the electroproduction of charm at various $Q^2$, with $W^{0.88}$ and pQCD fits (upper and lower curves, respectively)}
\label{CHARM}
\vskip 7truemm
\bc
\epsfxsize=0.69\hsize\epsfbox[75 560 350 765]{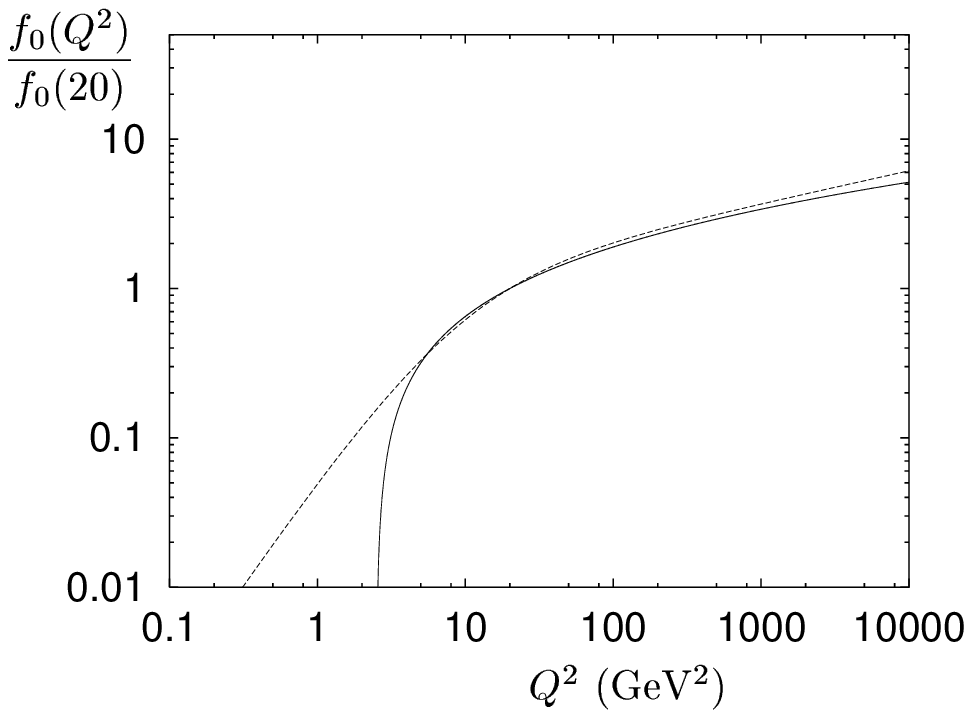}
\ec
\vskip -12truemm
\caption{NLO evolution of the hard-pomeron coefficient function (solid curve)
with the phenomenological fit (broken curve)}
\label{evol}
\end{figure}

Data\cite{zeusc} for the charm structure function $F_2^c(x,Q^2)$ have the
remarkable property that, at all available $Q^2$, they fit to just
the single hard-pomeron power of $x$. Further, to an excellent approximation
the coupling of the hard pomeron appears to be flavour blind:
\be
F_2^c(x,Q^2)=f_c(Q^2)\,x^{-\epsilon_0}
\ee
with
\be
f_c(Q^2)={0.4}~f_0(Q^2)
\ee
So if we define a charm-production cross section
\be
\sigma^c(W)={4\pi^2\alpha_{\hbox{{\fiverm EM}}}\over Q^2}F_2^c(x,Q^2)
\Big|_{x=Q^2/(W^2+Q^2)}
\ee
it behaves as $W^{2\epsilon_0}$ at all $Q^2$, even down to \hbox{$Q^2=0$}: see
figure \ref{CHARM}.
Perturbative QCD directly relates $F_2^c(x,Q^2)$
to the gluon structure function, so that at small $x$ it too must be 
dominated by hard-pomeron exchange alone, even at quite small values of
$Q^2$. This what causes the rapid rise  at small $x$ of the
DL curve in figure \ref{GLUONSF}.

\def\u{{\bf u}}\def\P{{\bf P}}\def\f{{\bf f}}
\section{DGLAP evolution}

The singlet DGLAP equation introduces the two-component
quantity
\be
\u(x,t)= \left (\matrix{x\sum _f(q_f+\bar q_f)\cr xg(x,t)\cr}\right )
\ee
where $t=\log (Q^2/\Lambda^2)$. If we Mellin transform with respect to
$x$, the equation becomes very simple:
\be
{{\pd\over\pd t}\u (N,Q^2)= \P (N,\alpha_s(Q^2))\,\u(N,Q^2)}
\label{dglap}
\ee
The usual approach is to expand the splitting matrix $\P$ in powers
of $\alpha_s$. However, this is mathematically illegal when $N$ is small.
Compare
\be
\sqrt{N^2+\alpha_s}-N=\alpha_s/2N-\alpha_s^2/8N^3+\dots
\ee
Here, each term in the expansion is singular at $N=0$ but the function
itself is regular there: the expansion is illegal for $N^2\leq \alpha_s$.
Similarly, it is likely that whenever expanding $\P (N,\alpha_s(Q^2))$
makes it large, and therefore makes $\u(N,Q^2)$ vary rapidly with
$Q^2$, the expansion is dangerous.

My own belief is that $\P (N,\alpha_s(Q^2))$ has no singularities in
the complex-$N$ plane, or at least no relevant singularities. My reason
is that solving (\ref{dglap}) would cause a singularity of
$\P (N,\alpha_s(Q^2))$ to induce an {\it essential} singularity 
in $\u(N,Q^2)$ (that is, a nasty one). The variable $N$ is closely
related to the orbital angular momentum $\ell$, and I was brought up\cite{elop} 
to believe that matrix elements such as $\u(N,Q^2)$ do not have essential 
singularities in the complex $\ell$-plane. This point of view contrasts
with  that of those who believe that the value of $\epsilon_0$ is associated
with a singularity of $\P (N,\alpha_s(Q^2))$ and may even be calculated,
perhaps by refining the BFKL approach. I think that very probably
$\epsilon_0$ is a nonperturbative quantity that therefore cannot be
calculated.

A fixed-power behaviour $x^{-\epsilon_0}$ of $F_2(x,Q^2)$, such as
in (\ref{regge}), corresponds to an $N$-plane pole:
\be
\u(N,Q^2)\sim {\f(Q^2)\over N-\epsilon_0}~~~~~~~\f(Q^2)=\Big(\matrix{f_0(Q^2)\cr f_g(Q^2)\cr}\Big )
\ee
If we insert this into the DGLAP equation
(\ref{dglap}) and equate the residue of the pole on each side
of the equation, we find
\be
{{\pd\over\pd t}\f(Q^2)= \P (N=\epsilon_0,\alpha_s(Q^2))\, \f(Q^2)}
\label{evolution}
\ee
$\epsilon_0$ is far enough from 0 for the expansion 
of $\P (N=\epsilon_0,\alpha_s(Q^2))$ to be reasonably safe. So we may
easily use the DGLAP equation to calculate the evolution of the
hard-pomeron component of $F_2(x,Q^2)$. But this is not the case for
the soft-pomeron component, because $\epsilon_1\approx 0.08$ is too
close to 0.

\begin{figure*}[t]
\bc
\epsfxsize=0.35\hsize\epsfbox[75 560 350 765]{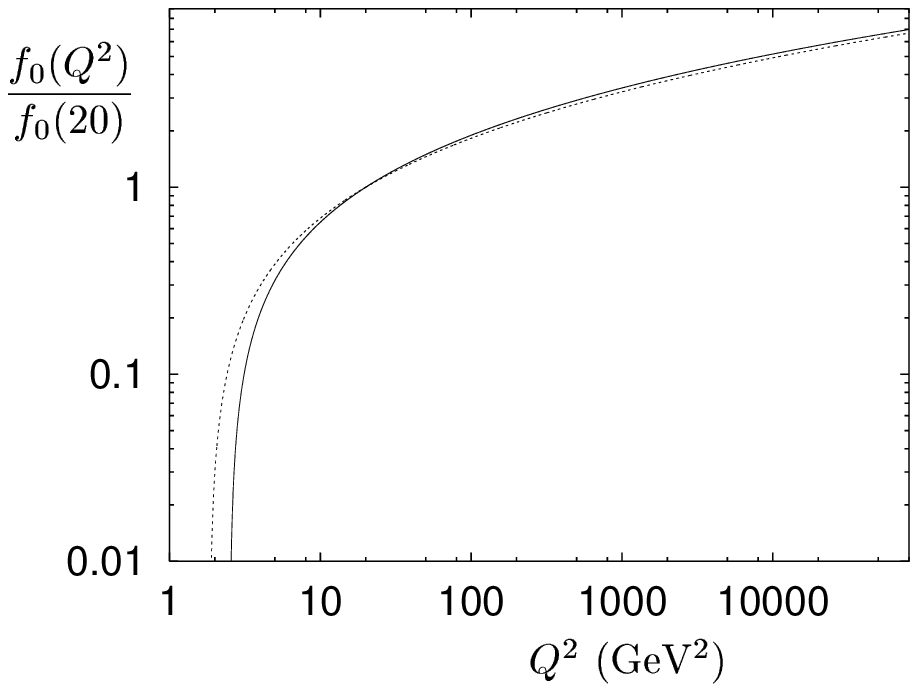}
\epsfxsize=0.37\hsize\epsfbox[60 560 345 760]{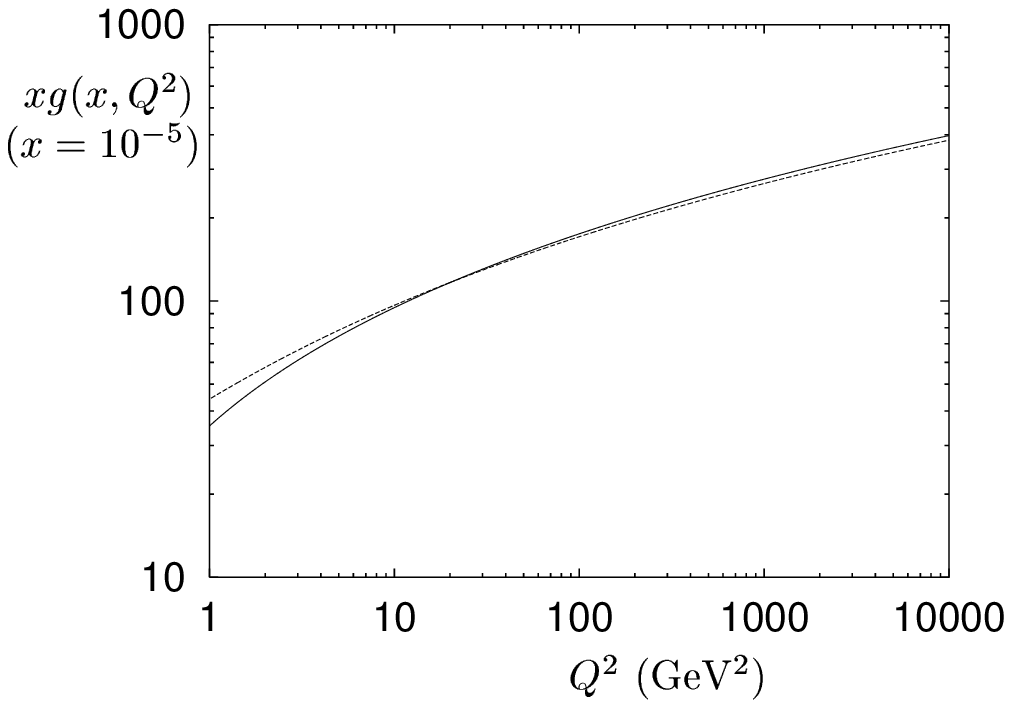}
\ec
\vskip -11truemm
\caption{LO and NLO evolution of the hard-pomeron and gluon coefficient
functions}
\label{leading}
\vskip 7truemm
\bc
\epsfxsize=0.36\hsize\epsfbox[50 50 390 290]{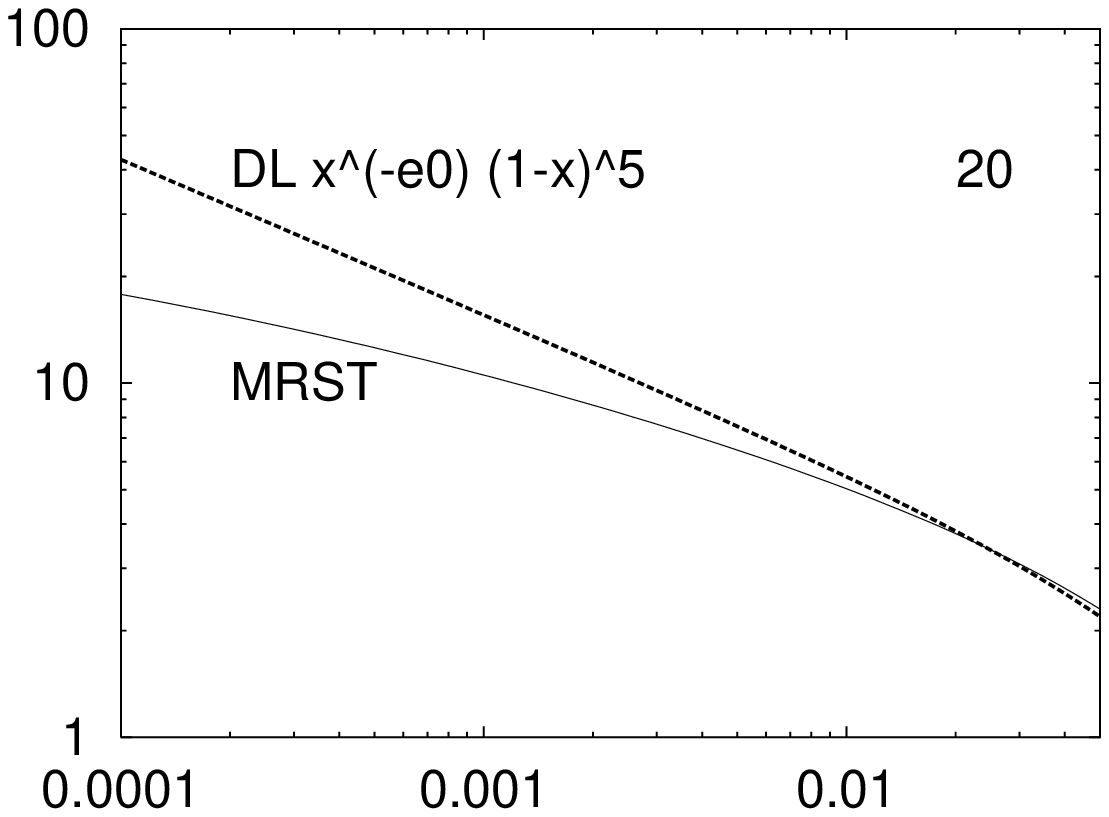}
\epsfxsize=0.36\hsize\epsfbox[50 50 390 290]{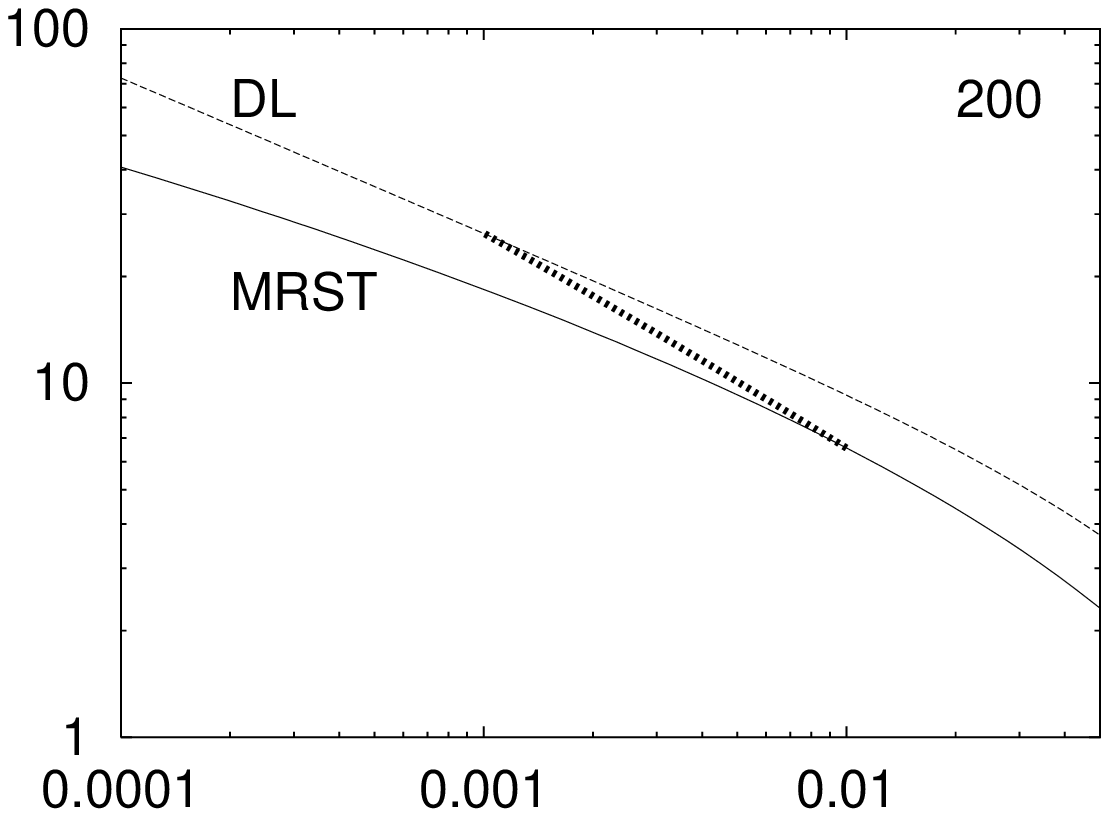}
\ec
\vskip -11truemm
\caption{Gluon structure function at $Q^2=20$ and 200 GeV$^2$}
\label{dlmrst}
\end{figure*}

According to figure \ref{GLUONSF}, the various gluon structure
functions come together at $x\approx 0.01$. It is reasonable to assume that
for values of $x$ larger than this the evolution of the two elements
of $\u(x,Q^2)$ does not use values of $N$ close to 0 and therefore the
conventional analysis is correct. 
So we can start at some not-too-large value of $Q^2$, 20 GeV$^2$ say.
We determine the value of $f_0(Q^2)$ there from the phenomenological fit
(\ref{pheno}) and $f_g(Q^2)$ from the MRST gluon structure function $xg(x,Q^2)$,
which for $x$ greater than about 0.01 fits very well to $x^{-\epsilon_0}
(1-x)^5$. 
We  choose $\Lambda_{\hbox{{\sevenrm NLO}}}$ such that
$\alpha_s(M_X^2)=0.116$ and use (\ref{evolution}) to calculate\cite{dl} the NLO
evolution of 
$f_0(Q^2)$ and $f_g(Q^2)$ in both directions. The result for $f_0(Q^2)$ is the 
continuous curve in figure \ref{evol}. The dashed curve is the 
phenomenological form (\ref{pheno}). Provided we adjust $\Lambda$
so that still $\alpha_s(M_X^2)=0.116$, LO evolution gives almost identical
results; this is shown in figure \ref{leading}.

The agreement between the pQCD calculation and the phenomenological
curve is a success not only for the concept of the hard pomeron, but
also for pQCD itself. The evolution is from a single value of $Q^2$,
not the customary global fit\cite{mrst,cteq}, and it introduces far fewer
parameters. 

Notice that, as $Q^2$ increases the large-$x$ behaviour of $xg(x,Q^2)$
becomes steadily steeper than $(1-x)^5$, and so the largest value of $x$
for which $x^{-\epsilon_0}$ is a good approximation to the structure
function steadily decreases. Figure \ref{dlmrst} shows an estimate
of this.

We may use the gluon structure function to calculate the charm structure
function $F_2^c(x,Q^2)$. The result, using just LO photon-gluon fusion
with a charm-quark mass $m_c=1.3$ GeV,
is the solid curves in figure \ref{CHARM}. This is an important check
on the consistency of the approach. As is seen in figure \ref{mrst1.8},
a steep gluon distribution is needed to fit the data at small $Q^2$.

In conclusion,
the conventional approach to evolution needs modifying at small $x$.
It can be corrected if we combine it with Regge theory,
but only partly --- we can only treat the hard-pomeron part.
The resulting gluon distribution
is larger at small $x$ than has so far been supposed
and gives a good description of charm production.
I should add that we want good data for the longitudinal structure function,
because this gives the most direct window on the gluon distribution.

\vskip -7truemm
\begin{figure}[h]
\bc
\epsfxsize=0.6\hsize\epsfbox[80 70 400 285]{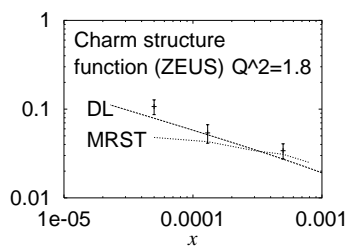}
\ec
\vskip -10truemm
\caption{Charm structure functions at $Q^2=1.8$ GeV$^2$, with ZEUS data\cite{zeusc}}
\label{mrst1.8}
\vskip -7truemm
\end{figure}

\end{document}